\documentclass[11pt,aps,prd,onecolumn,nofootinbib,amsmath,amssymb,longbibliography,floatfix]{revtex4-2}

\usepackage{amsbsy,amsfonts}
\usepackage{graphicx}
\graphicspath{{figures/}}
\usepackage{booktabs}
\usepackage{array}
\usepackage{placeins}
\usepackage{slashed}
\usepackage[colorlinks,citecolor=blue,linktoc=all,linkcolor=cyan,urlcolor=blue]{hyperref}

\usepackage{setspace}
\newcommand{\dd}{\mathrm{d}}
\newcommand{\impart}{\operatorname{Im}}

\newcommand{\beq}{\begin{equation}}
\newcommand{\eeq}{\end{equation}}
\newcommand{\bea}{\begin{eqnarray}}
\newcommand{\eea}{\end{eqnarray}}

\begin{document}
\raggedbottom

\title{Tensor-structure sum rules for spin-1 targets at all \texorpdfstring{$Q^2$}{Q2}}

\author{Vladimir Pascalutsa}
\author{Marc Vanderhaeghen}
\affiliation{Institut f\"ur Kernphysik, Johannes Gutenberg-Universit\"at Mainz,
D-55128 Mainz, Germany}

\date{\today}

\begin{abstract}
We derive a family of sum rules for the tensor structure functions of targets with spin one or higher, rooted in exact Siegert relations.  The mixed rank-two $TT$--$LT$ degeneracy combines amplitudes of opposite crossing parity and, under the superconvergence assumption, gives
$\int_0^1\!\dd x\,[b_2(x,Q^2)-3b_4(x,Q^2)]=0$
in the Hoodbhoy--Jaffe--Manohar (HJM) basis.
At large $Q^2$, the protected moment agrees with Detmold's target-mass-complete twist-two OPE; in the strict Bjorken projection, and if $b_4$ is neglected, it reduces to the Efremov--Teryaev (ET)-type condition $\int_0^1\!\dd x\,x\,b_1(x)=0$, whose measured-range footprint is consistent with the HERMES data.  Two further rank-two combinations connect crossing-even amplitudes and determine the corresponding subtraction functions through weighted HJM moments.  The protected combination can be isolated through separated tensor $LT$ and $TT$ responses, providing a direct experimental strategy for JLab and, ultimately, a future EIC.
\end{abstract}

\maketitle

\section{Introduction}
\label{sec:Lintro}

Sum rules turn inclusive observables into statements about structure that is otherwise hard to isolate.  In deep-inelastic scattering (DIS), the partonic sum rules (e.g., Bjorken~\cite{Bjorken:1966jh}, Gottfried~\cite{Gottfried:1967kk}, Ellis--Jaffe~\cite{Ellis:1973kp}) have been as instructive in their failures as in their successes: the Gottfried defect exposed a flavor-asymmetric sea~\cite{NewMuon:1991hlj}, while the Ellis--Jaffe shortfall opened the proton-spin problem~\cite{EuropeanMuon:1987isl}.  These relations hold within dynamical approximations such as the naive parton model or leading twist; when they fail, the failure itself carries information about the dynamics beyond them.

A different class of relations begins with general constraints rather than a constituent picture.  The Burkhardt--Cottingham (BC) relation for $g_2$~\cite{Burkhardt:1970ti}, the Gerasimov--Drell--Hearn relation~\cite{Gerasimov:1965et,Drell:1966jv}, and related forward-Compton sum rules~\cite{Drechsel:2002ar,Hagelstein:2015egb,Pasquini:2018wbl,Pascalutsa:2024hvy} follow from causality, unitarity, crossing, and electromagnetic gauge invariance.  They are model independent apart from assumptions about analytic structure and convergence.  In this work, we discuss whether the tensor sector of a massive spin-1 target admits a relation of this kind.

\begin{figure}[hbt]
\centering
\includegraphics[width=0.9\textwidth]{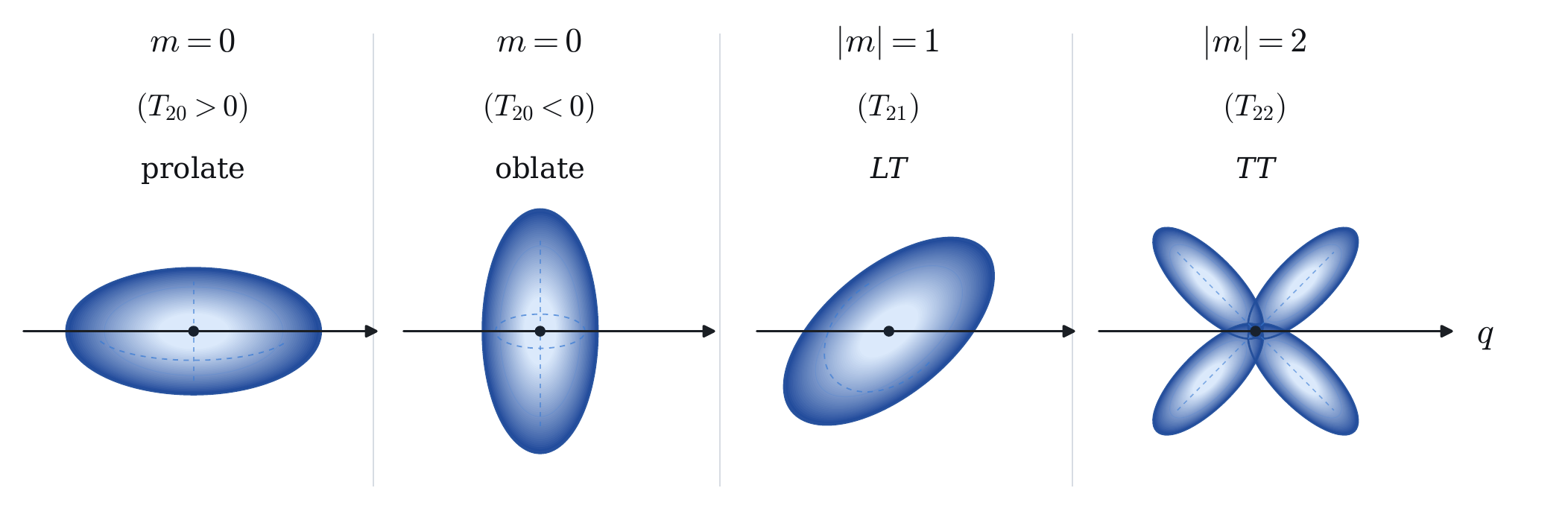}
\vspace{-0.7em}
\caption{Characteristic rank-two alignment patterns relative to the photon momentum $\boldsymbol q$.  The shapes are symbolic spin-tensor representations, not spatial or charge distributions.}
\label{fig:introAlignment}
\end{figure}

A target of spin one or higher may also carry tensor polarization, or \textit{alignment}: the quadrupole component of its spin distribution relative to a chosen axis.  A spin-$s$ density matrix decomposes into irreducible moments of ranks $k=0,\ldots,2s$~\cite{Fano:1957zz}.  Spin-one therefore contains five rank-two polarization components~\cite{Ohlsen:1972zz,Bacchetta:2000jk}, but not five independent electromagnetic responses.  Once the photon direction is chosen, axial symmetry, parity, and current conservation leave four rank-two helicity responses: two with zero helicity transfer and one each with $|m|=1$ and $|m|=2$, as illustrated in Fig.~\ref{fig:introAlignment}.  They are conventionally parametrized by the tensor structure functions $b_{1,\ldots,4}$ introduced by Hoodbhoy, Jaffe, and Manohar (HJM)~\cite{Hoodbhoy:1988am}.

In the naive partonic interpretation, the leading-twist function $b_1$ measures the difference between quark distributions in the $m=0$ and $|m|=1$ target states.  Its best-known constraint is the Close--Kumano sum rule which predicts a vanishing unweighted moment~\cite{Mankiewicz:1988dk,Close:1990zw}.  Related tensor-polarization constraints were derived by Efremov and Teryaev (ET)~\cite{Efremov:1981vs} and later translated into the HJM language~\cite{Umnikov:1996qv,Efremov:1994xf}. Their momentum-weighted relation reappears below as a footprint of an exact
BC-type superconvergence in the Bjorken limit.

The deuteron is where these rather abstract rank-two structures become concrete. At nuclear resolution, the tensor force and the 
$D$-state make the
spin-projected density different for $m=0$ and $|m|=1$; these are conditional density
contours, not rigid intrinsic shapes~\cite{Forest:1996kp}. Relativistic binding, coherent double scattering, hidden-color components, and tensor-polarized sea quarks then provide physically distinct ways of generating a nonzero tensor response
~\cite{Edelmann:1997qe,Edelmann:1997ik,Nikolaev:1996jy,Bora:1997pi,Kumano:2010vz,Miller:2013hla}.

The experimental record is still thin but has a promising future. HERMES remains the only DIS measurement of $b_1$ and found the unweighted moment different from zero~\cite{HERMES:2005pon,Riedl:2005jq}. Jefferson Lab experiments E12-13-011 and
E12-15-005 address $b_1$ and tensor asymmetries with a polarized deuteron target
~\cite{DeurB1Proposal:2013,JLab:PAC51,Poudel:2025nof,Cosyn:2024drt}, while developments in tensor-polarized
targets open further possibilities~\cite{Dalton:2025jzt}. A future
Electron--Ion Collider (EIC) would extend the reach toward small $x$, tensor-polarized antiquarks, and gluon transversity, which has no spin-$1/2$ analogue
~\cite{AbdulKhalek:2021gbh,Kumano:2024fpr,Kumano:2021fem}. Tensor structure is therefore beginning to move from a mostly formal subject toward one in which different moments and response functions can be confronted separately.

The partonic sum rules are only part of this story.  To confront finite-$Q^2$ experiments, sum rules of the BC type, whose content does not depend on a particular constituent picture, are of general  interest to test.  The derivation below uses only the mass and spin of the target; QCD or nuclear dynamics enter only when the resulting relations are evaluated in a particular system.

The paper is organized as follows. Section~\ref{sec:Lsiegert} derives the three rank-two
sum rules together and then singles out the mixed $TT$--$LT$ relation as the superconvergent one. 
The elastic and
low-energy limits are discussed in Section~\ref{sec:Lelasticlow}. 
Section~\ref{sec:Lope} is devoted to their large-$Q^2$
content: the Bjorken/ET footprint, Detmold's inclusion of target-mass corrections and the resulting twist-two relations, and
the asymptotic behavior of the two tensor subtraction functions. 
Section~\ref{sec:Lmeasuresec} discusses the present experimental situation, derives the direct
projection needed for a finite-$Q^2$ test, applying it to the HERMES data as an illustration of the Bjorken-limit footprint. We then summarize the implications in the Conclusion.  The conventions and the dictionaries among the HJM, Budnev, and Edelmann--Piller--Weise bases and the complete elastic structure functions are collected in the Supplementary
Material.

\section{Three tensor sum rules}
\label{sec:Lsiegert}

The construction needs no model of the target.  It starts from analyticity, unitarity, parity, and time-reversal invariance, applied to the spin-1 forward doubly-virtual Compton scattering (VVCS) amplitude.  With photon and target four-momenta $q$ and $p$, the relevant invariants are
\beq
\label{eq:kin}
    q^2=-Q^2, \quad p^2 =M^2, \quad \nu=\frac{p\!\cdot q}{M}.
\eeq
The photon energy in the lab frame $\nu$ can be traded for the Bjorken variable $x=Q^2/(2M\nu)$. 

The key kinematic point is the Siegert point, $\nu_S=\pm iQ$,
where the virtual photon has zero three-momentum in the target rest frame.
The longitudinal ($L$) and transverse ($T$) responses then become degenerate.  We refer to this as the \emph{Siegert LT-degeneracy}~\cite{Pascalutsa:2025suw}.  In broader terms, it is a direct manifestation of rotational symmetry: with no spatial photon momentum, longitudinal and transverse states are different projections of the same rotational multipoles.  As a result,
unsubtracted dispersion relations evaluated at that point lead to
BC-type superconvergence, whereas once-subtracted relations lead to sum rules for the subtraction functions.

The superconvergent channel is already visible in the following pair of rank-two helicity amplitudes~\footnote{We denote the helicity amplitudes of the VVCS amplitudes as
${\cal M}_{\lambda'_\gamma  \lambda'_D, \lambda_\gamma \lambda_D}$, with $\lambda_\gamma (\lambda'_\gamma)$ the helicities of initial (final) virtual photons, and $\lambda_D (\lambda'_D)$ the helicities of initial (final) deuterons respectively, where we specify the 
conventions used in this paper in Appendix~\ref{app:Budnevbasis}. 
}:
\beq
 {\cal M}_{TT}\equiv{\cal M}_{++,--},\qquad
 {\cal M}_{LT}\equiv{\cal M}_{++,00}+{\cal M}_{0+,-0}.
 \label{eq:Lhelicity}
\eeq
Crossing makes the first amplitude even and the second odd in $\nu$,
${\cal M}_{TT}(-\nu,Q^2)={\cal M}_{TT}(\nu,Q^2)$ and
${\cal M}_{LT}(-\nu,Q^2)=-{\cal M}_{LT}(\nu,Q^2)$.  If their high-energy
behavior permits unsubtracted dispersion relations, they satisfy 
\begin{subequations}
\label{eq:LdispRelations}
\bea
 {\cal M}_{TT}(\nu,Q^2)
  & = & \frac{2}{\pi}\int_{\nu_{\rm el}}^\infty\dd\nu'\,
 \frac{\nu' W_{TT}(\nu',Q^2)}{\nu'^2-\nu^2},
 \label{eq:LdispTT}\\
 {\cal M}_{LT}(\nu,Q^2)
  & = & \frac{2\nu}{\pi}\int_{\nu_{\rm el}}^\infty\dd\nu'\,
 \frac{W_{LT}(\nu',Q^2)}{\nu'^2-\nu^2},
 \label{eq:LdispLT}
\eea
\label{eq:Ldisp}
\end{subequations}
where $W_{TT}=\impart {\cal M}_{TT}$ and $W_{LT}=\impart {\cal M}_{LT}$ on the physical discontinuity.  The elastic pole lies at $\nu_{\rm el}=Q^2/(2M)$, while the inelastic cut begins at $\nu_{\rm th}$.  In the HJM basis, the corresponding absorptive responses are (cf.~Appendix~\ref{app:dictionary}):
\begin{subequations}\label{eq:Lresponses}
\bea
 W_{TT} & = & \frac{M}{\nu}\left(\frac13b_2-b_3\right),
 \label{eq:LresponseTT}\\
 W_{LT} & = & \frac{M}{Q}\left[\frac13b_2-b_3
 +\frac16v_\gamma^2(b_2-3b_4)\right].
 \label{eq:LresponseLT}
\eea
\label{eq:Lresponse}
\end{subequations}

At the Siegert point the Wigner--Eckart theorem relates both structure functions, with the standard analytic-continuation phase, as
\beq
 \nu_S{\cal M}_{TT}(\nu_S,Q^2)=Q{\cal M}_{LT}(\nu_S,Q^2).
 \label{eq:Lsiegert}
\eeq
This is an exact local statement for $\nu = \nu_S $.  On the physical absorptive support, the same combination becomes
\beq
 \nu W_{TT}-QW_{LT}
 =-\frac{M}{6}v_\gamma^2\,[b_2-3b_4],
 \label{eq:WTTminusTL}
\eeq
with $v_\gamma^2 \equiv (\nu^2 + Q^2)/\nu^2 $. 
Inserting Eqs.~\eqref{eq:Lsiegert} and \eqref{eq:WTTminusTL} into the two dispersion relations and expressing them as integral over $x$, at any value of $Q^2$, gives
\beq
 \int_0^1\dd x\,[b_2(x,Q^2)-3b_4(x,Q^2)]=0 .
 \label{eq:Lsumrule}
\eeq
This is the BC-type tensor superconvergence relation.  Its logic is worth stressing: the local Siegert identity of Eq.~(\ref{eq:Lsiegert}) is exact, while the vanishing moment additionally assumes the high-energy convergence needed for the unsubtracted dispersion relation in the combination entering Eq.~\eqref{eq:Lsumrule} to hold.  In the helicity basis of Edelmann, Piller, and Weise~\cite{Edelmann:1997qe,Edelmann:1997ik}, the integrand $b_2-3b_4$ combines the double-helicity-flip response with the mixed tensor response, see Eq.~\eqref{eq:EdelmannInvProtected} in Appendix~\ref{app:Edelmann}; at finite $Q^2$ the sum rule is therefore not a constraint on double-helicity-flip absorption alone.  Should the superconvergence hypothesis fail, e.g., through a fixed pole in the crossing-odd combination, the right-hand side of Eq.~\eqref{eq:Lsumrule} would acquire a nonvanishing $Q^2$-dependent term; we return to this possibility in Secs.~\ref{sec:Lope} and \ref{sec:Lconclusion}.

The same rank-two Wigner--Eckart reduction contains two further independent relations.  We define the corresponding combinations by
\begin{subequations}\label{eq:LdiagonalPhidefs}
\bea
 {\cal M}_{T\text{--}\Delta}(\nu,Q^2) & \equiv & 
 {\cal M}_{++,++}+{\cal M}_{+-,+-}-2{\cal M}_{+0,+0}-{\cal M}_{++,--},
 \label{eq:LPhiTdef}\\
 {\cal M}_{L\text{--}\Delta}(\nu,Q^2) & \equiv & 
 {\cal M}_{0+,0+}-{\cal M}_{00,00}-{\cal M}_{++,--}.
 \label{eq:LPhiLdef}
\eea
\end{subequations}
where the subscript $\Delta$ indicates the presence of the double-spin-flip amplitude ${\cal M}_{++,--}$ in the corresponding difference.

These two linear combinations vanish at the Siegert point,
\beq
 {\cal M}_{T-\Delta}(iQ,Q^2)={\cal M}_{L-\Delta}(iQ,Q^2)=0,
 \label{eq:LPhiSiegertZero}
\eeq
which again is a direct consequence of rotational invariance.  At this point $q^\mu=(iQ,\boldsymbol 0)$, current conservation leaves only
the three spatial current components, while the analytically continued longitudinal photon
polarization becomes the third spatial spherical component (up to its standard phase).
With no spatial momentum left to select an axis, the symmetric traceless part of the
spatial current product is a single rank-two irreducible tensor.  The three rank-two helicity projections entering Eq.~\eqref{eq:LdiagonalPhidefs} are therefore just different
Clebsch--Gordan projections of one reduced matrix element.  In the normalization of
Eq.~\eqref{eq:LdiagonalPhidefs}, the $T$ and $L$ projections equal the $\Delta$ projection; the analytic-continuation phase of the longitudinal polarization fixes the
relative sign.

The same statement can be seen algebraically in the HJM basis.  On the physical absorptive support, denoting the
corresponding absorptive differences by $W_{T-\Delta}$ and $W_{L-\Delta}$, one has
\bea
 W_{T-\Delta}
 & = & -2v_\gamma^2 b_1,\nonumber\\
 W_{L-\Delta}
 & = & v_\gamma^2\left[
 b_1-\frac{M\nu}{Q^2}b_2
 -\frac{M}{\nu}\left(\frac13b_2+b_3+b_4\right)\right].
\label{eq:LdiagonalW}
\eea
Like the mixed combination in Eq.~\eqref{eq:WTTminusTL}, both differences carry the same
kinematic factor $v_\gamma^2$.  The common factor is the clue: all three rank-two combinations already know about the same rotational symmetry at the level of the discontinuity. Their zeros are not accidental cancellations among the $b_i$.

For these two crossing-even amplitudes, we use a once-subtracted dispersive representation, of the form 
\bea
 {\cal M}_{T-\Delta}(\nu,Q^2) 
  & = & {\cal M}_{T-\Delta}(0,Q^2) + \frac{2}{\pi} \nu^2 \int_{\nu_{\rm el}}^\infty\dd\nu'\,
 \frac{ W_{T-\Delta}(\nu',Q^2)}{\nu' (\nu'^2-\nu^2)},
 \label{eq:dispT-Delta}
\eea
and an analogous relation for ${\cal M}_{L-\Delta}(\nu,Q^2)$.  
Evaluating it at the Siegert point $\nu = i Q$ gives the two companion
sum rules
\begin{subequations}\label{eq:LtensorSubtractions}
\bea
 {\cal M}_{T-\Delta}(0,Q^2)
  & = & -\frac{16M^2}{\pi Q^2}\int_0^1\dd x\,x\,b_1(x,Q^2),
 \label{eq:LPhiTsum}\\
 {\cal M}_{L-\Delta}(0,Q^2)
  & = & \frac{8M^2}{\pi Q^2}\int_0^1\dd x\left[x\,b_1-\frac12 b_2
 -\frac{2M^2x^2}{Q^2}\left(\frac13b_2+b_3+b_4\right)\right](x,Q^2).
 \label{eq:LPhiLsum}
\eea
\end{subequations}
The integrals in Eqs.~\eqref{eq:LPhiTsum} and \eqref{eq:LPhiLsum} are exact finite-$Q^2$ moments of the physical structure functions.  In particular, the moment in Eq.~\eqref{eq:LPhiTsum} should not be identified with the asymptotic OPE moment $\Gamma_2[b_1]$ introduced in Sec.~\ref{sec:Lope}.  The latter appears only after the twist-two, large-$Q^2$ expansion of the finite-$Q^2$ moment is performed.

Subtraction-function sum rules of this kind are familiar
in unpolarized VVCS~\cite{Gasser:2015dwa}; the resulting zero-energy subtraction functions are the
rank-two analogue of the $T_1(0,Q^2)$ construction of Ref.~\cite{Biloshytskyi:2023fyv}.

The broader pattern is summarized in Table~\ref{tab:SiegertClassification}.  Together with the unpolarized and vector-polarized cases, the three rank-two relations above complete the five local relations expected when the eight spin-one structure functions collapse at the Siegert point to three rotationally reduced structures proportional to $F_1$, $\frac13b_2-b_3$, and $g_1+g_2$:
\begin{table}[h]
\centering
\small
\begin{tabular}{@{}llll@{}}
\toprule
sector & helicity pair & crossing & dispersive consequence \\
\midrule
rank 0 & unpolarized $L-T$ & even/even & subtraction function $T_1(0,Q^2)$ \\
rank 1 & spin $TT-TL$ & odd/even & BC sum rule $\int \dd x\,g_2=0$ \\
rank 2 & tensor $TT-TL$ & even/odd & $\int \dd x\,(b_2-3b_4)=0$ \\
rank 2 & $T-\Delta$ & even/even & ${\cal M}_{T-\Delta}(0,Q^2)\leftrightarrow\int \dd x\,x b_1$ \\
rank 2 & $L-\Delta$ & even/even & ${\cal M}_{L-\Delta}(0,Q^2)$ weighted moment \\
\bottomrule
\end{tabular}
\caption{Forward spin-one degeneracies at the Siegert point, organized by tensor rank and crossing parity.  Opposite crossing parity produces the BC and tensor superconvergent sum rules.  The same-parity $T-\Delta$ and $L-\Delta$ relations instead determine once-subtracted Compton amplitudes.}
\label{tab:SiegertClassification}
\end{table}

\section{Elastic--inelastic balance}
\label{sec:Lelasticlow}

A sum rule involving the full structure functions must include the elastic pole.  The elastic electromagnetic structure of a spin-1 target is characterized by the charge, magnetic, and quadrupole form factors $G_C$, $G_M$, and $G_Q$ respectively. The elastic contribution to the integrand of Eq.~(\ref{eq:Lsumrule}) is given by
\beq
 [b_2-3b_4]^{\rm el}
 =-6\tau^2G_M(Q^2)[G_M(Q^2)+2G_Q(Q^2)]\,\delta(1-x),
 \label{eq:Lelastic}
\eeq
with $\tau=Q^2/(4M^2)$. 
One notices from Eq.~(\ref{eq:Lelastic}) that in this combination the charge form factor drops out.  

The other two sum rules also have definite elastic pieces.  Inserting the elastic HJM functions collected in Appendix~\ref{app:elastic} into Eqs.~\eqref{eq:LPhiTsum} and \eqref{eq:LPhiLsum} gives
\begin{subequations}\label{eq:LsubElastic}
\bea
 {\cal M}_{T-\Delta}^{\rm el}(0,Q^2)
  & = & -\frac{2}{\pi}\,G_M^2(Q^2),
 \label{eq:LsubElasticT}\\
 {\cal M}_{L-\Delta}^{\rm el}(0,Q^2)
  & = & -\frac{1}{\pi}\left[G_M^2(Q^2)+4G_C(Q^2)G_Q(Q^2)
  +\frac43\tau G_Q^2(Q^2)\right].
 \label{eq:LsubElasticL}
\eea
\end{subequations}
The latter are not elastic-inelastic cancellation conditions.  The $T-\Delta$ and $L-\Delta$ sum rules determine nonzero subtraction functions, so their elastic terms simply add to the corresponding inelastic moments.  At the photon point, with $G_C(0)=1$, Eqs.~\eqref{eq:LsubElasticT} and \eqref{eq:LsubElasticL} approach $-2G_M^2(0)/\pi$ and $-[G_M^2(0)+4G_Q(0)]/\pi$, respectively.  At the natural spin-1 values $G_M(0)=2$, $G_Q(0)=-1$, the longitudinal elastic subtraction term vanishes whereas the transverse one remains $-8/\pi$.  The protected pole in Eq.~\eqref{eq:Lelastic} is kinematically $O(Q^4)$; for a regular elastic vertex its coefficient is proportional to $G_M(0)[G_M(0)+2G_Q(0)]$.  It therefore also vanishes at the natural values, and the protected elastic pole starts only at $O(Q^6)$.

All three elastic contributions become rapidly irrelevant at large $Q^2$.  Perturbative-QCD helicity conservation for a composite spin-1 system gives $G_C:G_M:G_Q=(1-\frac23\tau):2:-1$~\cite{BrodskyHiller:1992}.  For a $q\bar q$ vector meson, $G_C\sim Q^{-2}$~\cite{LepageBrodsky:1980} then implies $G_M,G_Q\sim Q^{-4}$. Before the leading $G_M+2G_Q=0$ cancellation, the protected elastic term scales as $Q^{-4}$, while the $T-\Delta$ and $L-\Delta$ elastic pieces scale as $Q^{-8}$ and $Q^{-6}$ respectively.  Thus, at large $Q^2$ the elastic endpoint is parametrically negligible compared with the inelastic structure-function contribution.\footnote{The deuteron large-$Q^2$ counting is even steeper.  Combining the spin-1 helicity ratio $G_C:G_M:G_Q=(1-\frac23\tau):2:-1$~\cite{BrodskyHiller:1992} with the constituent-counting behavior $G_C\sim Q^{-10}$~\cite{BrodskyJiLepage:1983} gives $G_M,G_Q\sim Q^{-12}$, up to logarithmic evolution.  Before the leading protected cancellation, Eq.~\eqref{eq:Lelastic} therefore scales as $Q^{-20}$, while Eqs.~\eqref{eq:LsubElasticT} and \eqref{eq:LsubElasticL} scale as $Q^{-24}$ and $Q^{-22}$, respectively.
}

For the protected relation of Eq.~\eqref{eq:Lsumrule} at low virtuality, superconvergence turns the statement into an elastic--inelastic balance. Converting the sum rule integral over $x$ into an integral over $\nu$ using the Budnev response dictionary of Appendix~\ref{app:dictionary}, one obtains
\beq
\begin{aligned}
 \int_0^{x_{\mathrm{th}}}\dd x\,[b_2-3b_4]
 &=-\frac{6Q^2}{M}\int_{\nu_{\mathrm{th}}(Q^2)}^\infty\dd\nu\,
 \frac{\nu\,\tau_{TT}(\nu,Q^2)-2Q\,\tau_{TL}(\nu,Q^2)}{\sqrt{\nu^2+Q^2}}\\
 &=6\tau^2G_M(Q^2)[G_M(Q^2)+2G_Q(Q^2)],
\end{aligned}
\label{eq:Linelbalance}
\eeq
with $x_{\mathrm{th}}=Q^2/(2M\nu_{\mathrm{th}})$. 
For a composite target there is no reason for this compensation to occur in a single channel.  Breakup, quasielastic strength, resonances, and the deep-inelastic continuum can all contribute.  For an elementary or narrow vector state, the same equation constrains whatever complete set of absorptive channels replaces them.  The sum rule does not require each contribution to be small; it constrains their signed total.

At the photon point the right-hand side of Eq.~\eqref{eq:Linelbalance} starts at $Q^4$, whereas the leading term on the dispersive side is proportional to $Q^2\int\dd\nu\,\tau_{TT}(\nu,0)$.  Their matching therefore gives the real-photon tensor superconvergence condition\footnote{The same condition noted long ago in  photon--photon fusion~\cite{Budnev:1973jv} is offset by a constant in leading-order QED verifications \cite{Worden:1974gh,Gerasimov:1973ja,Budnev:1980fs,Pascalutsa:2010sj}. The scalar/spinor QED tests come in ratio $+1:-2$, suggestive of $(-1)^{2s}(2s+1)$~\cite[Sec.~8.4]{Pascalutsa:2024hvy}.}
\beq
 \int_{\nu_{\mathrm{th}}}^\infty\dd\nu\,\tau_{TT}(\nu,0)=0.
 \label{eq:LrealSR}
\eeq
 with $\tau_{TT}=\sigma_{2\parallel}-\sigma_{2\perp}$ given in terms of cross sections linearly-polarized photoabsorption on transversely polarized target.  The absence of the $Q^2$ term is therefore the photon-point imprint of the same superconvergence assumption. 
 %; the next term in the expansion is collected in Appendix~\ref{sec:elastic}.
More generally, inverse-energy moments of $\tau_{TT}$ determine rank-two electric and magnetic dipole polarizabilities~\cite{Chen:1998vi,Ji:2003ia,Friar:2005gv}.  The next order of the low-$Q^2$ expansion of Eq.~\eqref{eq:Linelbalance} relates the tensor responses to $G_M(0)[G_M(0)+2G_Q(0)]$, i.e., to the magnetic and quadrupole moments, in analogy with Schwinger's sum rule for the anomalous magnetic moment~\cite{Hagelstein:2017obr}; for the deuteron this product is large, $\simeq 92$.  Similarly, the low-$Q^2$ expansion of the two subtraction functions yields rank-two analogues of the Bernab\'eu--Tarrach relation~\cite{Biloshytskyi:2023fyv} for the tensor polarizabilities.  We leave both to future work.  

\section{ET footprint and target mass corrections to leading-twist structure functions}
\label{sec:Lope}

We now focus on the large-$Q^2$ content of the three relations established above.  The limit gives a useful first impression, but it must be taken in the right order.  In the strict collinear leading-power projection, the physical tensor $LT$ and $TT$ responses are suppressed and the tensor Callan--Gross relation gives $b_2\to2x b_1$.  If this fixed-$x$ projection is made before the moment is formed, and if, in addition, $b_4$ is neglected, the protected moment has the following limit
\beq
 \int_0^1\dd x\,[b_2-3b_4](x, Q^2)
 \;\xrightarrow[\text{collinear}]{Q^2\to\infty}\;
 2\int_0^1\dd x\,x\,b_1(x) =0,
 \label{eq:LET}
\eeq
which is the electromagnetic $b_1$ footprint of the singlet momentum constraint discussed by Efremov and Teryaev~\cite{Efremov:1994xf}; we refer to it as the `ET-type momentum sum rule'.  The underlying ET singlet relation expresses a quark--gluon momentum balance, whereas the electromagnetic $b_1$ moment carries the usual charge-squared flavor weights.  It should be distinguished from the more familiar Close--Kumano sum rule for the unweighted moment~\cite{Mankiewicz:1988dk,Close:1990zw}: 
\beq 
\int_0^1\!\dd x\,b_1(x)=0,
\label{eq:CK}
\eeq
which in the simple parton picture 
implies the absence of a tensor-polarized sea.  

If $b_1$ were symmetric under $x\leftrightarrow1-x$, i.e.  $b_1(x)=b_1(1-x)$, then the vanishing of either moment implies the vanishing of the other.  
No such symmetry is however implied, or required, by QCD: the small- and large-$x$ regions correspond to distinct dynamics.  In this case 
the ET-type weighted moment, being selected by the protected combination in the collinear limit, has a better chance for survival.  The HERMES data discussed in Sec.~\ref{sec:Lmeasuresec} make this distinction especially visible (cf.\ Fig.~\ref{fig:HERMES}).

Eq.~\eqref{eq:LET}, however, is a strict Bjorken limit result.  Here the order matters.  Taking the fixed-$x$ collinear projection before the $x$ integration gives the ET footprint, but it discards target-mass correction terms that survive when the exact leading-twist moment is formed first.  Detmold's operator-product expansion (OPE) analysis~\cite{Detmold:2005iz} supplies precisely this target-mass-complete statement, to which we turn next.

Defining 
\beq
 \Gamma_n[f]\equiv\int_0^1\dd x\,x^{n-1}f(x),
 \qquad (Q^2\to\infty),
 \label{eq:LGammaDefinition}
\eeq
for the asymptotic (massless-target) twist-two moments, where their logarithmic QCD scale dependence is left implicit.  This notation is deliberately distinct from the finite-$Q^2$ moments appearing in the exact subtraction relations above. The target-mass-complete twist-two relations have been worked out in~\cite{Detmold:2005iz} and can be expressed as
\begin{subequations}\label{eq:LDetmold}
\bea
 \int_0^1\dd x\,x^{n-1} b_3^{\rm tmc}(x,Q^2)
  & = & \frac{n^2+n-6}{3n(n+1)}
  \int_0^1\dd x\,x^{n-1} b_2^{\rm tmc}(x,Q^2),
 \label{eq:LDetmoldb3}\\
 \int_0^1\dd x\,x^{n-1} b_4^{\rm tmc}(x,Q^2)
  & = & \frac{n^2-11n+12}{3n(n+1)}
  \int_0^1\dd x\,x^{n-1} b_2^{\rm tmc}(x,Q^2),
 \label{eq:LDetmoldb4}
\eea
\end{subequations}
with $n=1,3,\ldots$, and the superscript denoting the twist-two contributions to the structure functions including their target-mass corrections (tmc), following the notation of Ref.~\cite{Yndurain:2006}.  Since these corrections are suppressed by powers of $M^2/Q^2$, they are sometimes referred to as kinematic higher twist, which is why we avoid a superscript referring to twist two alone.
The second relation at $n=1$ becomes
\beq
 \int_0^1\dd x\,b_4^{\rm tmc}(x,Q^2)
 =\mbox{$\frac13$}\int_0^1\dd x\,b_2^{\rm tmc}(x,Q^2),
 \label{eq:Lcombo}
\eeq
and hence verifies the tensor superconvergence relation~\eqref{eq:Lsumrule} within the target-mass-complete twist-two sector. Whether the agreement persists order-by-order in a twist expansion remains open.

We next apply Detmold's target-mass-complete twist-2 analysis to the 
subtraction functions of Eqs.~\eqref{eq:LPhiTsum} and \eqref{eq:LPhiLsum}. In the $L-\Delta$ subtraction function, one needs the  combination
\beq
 \int_0^1\dd x\,x^{n-1}
 \left[\mbox{$\frac13$} b_2+b_3+b_4\right]^{\rm tmc}(x,Q^2)
 =\frac{(n-1)(n-2)}{n(n+1)}
 \int_0^1\dd x\,x^{n-1}b_2^{\rm tmc}(x,Q^2),
 \label{eq:LLDeltaMoment}
\eeq
which for $n=3$ yields the coefficient $1/6$.  Using the asymptotic relation $b_2\to2x b_1$, the target-mass expansion may then be expressed entirely through the asymptotic moments $\Gamma_n[b_1]$ of Eq.~\eqref{eq:LGammaDefinition}.

It is important to stress that Eq.~\eqref{eq:LPhiTsum} itself remains an exact finite-$Q^2$ relation.  Only its target-mass-corrected twist-two component is expanded in asymptotic moments. 
Keeping the leading terms in this expansion yields\footnote{These coefficients hold at leading order in the coefficient functions and follow from the trace subtraction of the tensor-polarized twist-two matrix elements.  The corresponding expansion of the unweighted $b_2$ moment is $\frac12\int_0^1\dd x\,b_2^{\rm tmc}=\Gamma_2[b_1]+\frac{1}{12}\frac{M^2}{Q^2}\Gamma_4[b_1]+\frac{2}{75}\frac{M^4}{Q^4}\Gamma_6[b_1]+\ldots$, so that the coefficients of Eq.~\eqref{eq:xb1b2TMC} follow as $7/12-1/12=1/2$ and $34/75-2/75=32/75$.  Although $b_2=2x\,b_1$ holds for the asymptotic moments, the first two terms of the integrand of Eq.~\eqref{eq:LPhiLsum} therefore contribute at the same order $M^4/Q^4$ as its explicitly mass-suppressed third term.}
\begin{subequations}
\bea
\int_0^1\dd x\, x\, b_1^{\rm tmc}(x,Q^2) &=& \Gamma_2[b_1] + 
\frac{7}{12}\frac{M^2}{Q^2}\,\Gamma_4[b_1]  
+\frac{34}{75} \frac{M^4}{Q^4}\,\Gamma_6[b_1]
+ O\!\left(\frac{M^6}{Q^6}\right), 
\label{eq:xb1TMC}
\\
\int_0^1\dd x\,  \left[ x b_1 - \mbox{$\frac12$} b_2 \right]^{\rm tmc} (x,Q^2) &=& 
\frac{1}{2}\frac{M^2}{Q^2}\,\Gamma_4[b_1] 
+\frac{32}{75} \frac{M^4}{Q^4}\,\Gamma_6[b_1]
+ O\!\left(\frac{M^6}{Q^6}\right).
\label{eq:xb1b2TMC}
\eea
\end{subequations}
In combination with Eq.~(\ref{eq:LLDeltaMoment}) for $n = 3$, one then obtains the following expansions for the subtraction functions of Eqs.~\eqref{eq:LPhiTsum} and \eqref{eq:LPhiLsum}
\begin{subequations}\label{eq:LsubOPE}
\bea
 {\cal M}_{T-\Delta}^{\rm tmc}(0,Q^2)
  & = & -\frac{16}{\pi}\frac{M^2}{Q^2}\,\Gamma_2[b_1]
   -\frac{28}{3\pi}\frac{M^4}{Q^4}\,\Gamma_4[b_1]
   +O\!\left(\frac{M^6}{Q^6}\right),
 \label{eq:LsubOPET}\\
 {\cal M}_{L-\Delta}^{\rm tmc}(0,Q^2)
  & = & -\frac{4}{3\pi}\frac{M^4}{Q^4}\,\Gamma_4[b_1]
   +O\!\left(\frac{M^6}{Q^6}\right).
 \label{eq:LsubOPEL}
\eea
\end{subequations}

The first term in Eq.~\eqref{eq:LsubOPET} is governed by the asymptotic moment $\Gamma_2[b_1]$; this should not be read as an identification of the finite-$Q^2$ subtraction function with that moment.  The OPE by itself does not require $\Gamma_2[b_1]$ to vanish.  If the fixed-$x$ collinear projection can be interchanged with the moment without a small-$x$, endpoint, or non-dispersive boundary contribution, the dispersive superconvergence implies the additional asymptotic condition $\Gamma_2[b_1]=0$.  It then removes the leading $O(M^2/Q^2)$ twist-two contribution to 
${\cal M}_{T-\Delta}$.  

In the longitudinal combination of Eq.~\eqref{eq:LsubOPEL} the corresponding $O(M^2/Q^2)$ term cancels at leading order in the coefficient functions, pushing the first surviving contribution to $O(M^4/Q^4)$ and $\Gamma_4[b_1]$. Beyond leading order, the difference of the relevant coefficient functions can generate an $O(\alpha_s M^2/Q^2)$ contribution to ${\cal M}_{L-\Delta}$ proportional to the same lowest $b_1$ moment.  The improved large-$Q^2$ suppression of ${\cal M}_{L-\Delta}$ is an OPE statement and does not, by itself, imply improved fixed-$Q^2$ dispersive convergence.

For empirical studies it may be useful to present the OPE relations in the pointwise form, similar to Wandzura--Wilczek (WW) in the rank-one sector~\cite{Wandzura:1977qf}.  Decomposing the two Mellin coefficients into $1/n$ and $1/(n+1)$ pieces gives the pair of twist-two WW representatives
\begin{subequations}\label{eq:LWW}
\bea
 b_3^{\rm tmc}[b_2](x)
  & = & \frac13 b_2(x)
 -2\int_x^1\frac{\dd y}{y}\,b_2(y)
 +2x\int_x^1\frac{\dd y}{y^2}\,b_2(y),
 \label{eq:LWWb3}\\
 b_4^{\rm tmc}[b_2](x)
  & = & \frac13 b_2(x)
 +4\int_x^1\frac{\dd y}{y}\,b_2(y)
 -8x\int_x^1\frac{\dd y}{y^2}\,b_2(y).
 \label{eq:LWWb4}
\eea
\end{subequations}
These spin-one WW relations have a noticeably more elaborate structure than the familiar rank-one relation for $g_2$: each contains a local term together with two distinct nonlocal kernels, $\int_x^1\dd y\,b_2(y)/y$ and $x\int_x^1\dd y\,b_2(y)/y^2$.  This richer form follows directly from the simultaneous $1/n$ and $1/(n+1)$ pieces in the Mellin coefficients of Eq.~\eqref{eq:LDetmold}, and is characteristic of the tensor twist-two sector.
Furthermore, substituting $b_2(x)\to2x b_1(x)$ allows one
to express all the twist-two tensor structure in terms of $b_1$.

The message from the OPE is sharper than a consistency check.  At twist two, QCD singles out exactly the lowest moment selected by this degeneracy, with target-mass trace terms included and perturbative corrections preserving the relation~\cite{Detmold:2005iz}.  The other two tensor channels do not share this protection: their subtraction functions contain ordinary moments and need not vanish.  What remains beyond the reach of the OPE is the high-energy dispersive boundary condition.  The all-$Q^2$ superconvergent sum rule therefore still requires the mixed $TT$--$LT$ amplitude to admit an unsubtracted dispersion relation, with no fixed-pole contribution.  A possible source of such a contribution is the leading-twist gluon double-helicity-flip (gluon transversity) response~\cite{Jaffe:1989xy}, which enters $b_2-3b_4$ through the $\Delta$ term of Eq.~\eqref{eq:EdelmannInvProtected}, weighted by a moment that is not constrained by the local OPE.

\section{Experimental status and direct test}
\label{sec:Lmeasuresec}

Experimentally, tensor DIS remains largely unexplored.  HERMES provides the only existing DIS measurement of $b_1$, while dedicated tensor-polarized deuteron programs at Jefferson Lab are approved for future running~\cite{HERMES:2005pon,Riedl:2005jq,Poudel:2025nof,Cosyn:2024drt,DeurB1Proposal:2013,Dalton:2025jzt}.  Already within the measured HERMES range, the two large-$Q^2$ moments discussed in Sec.~\ref{sec:Lope} behave strikingly differently: the unweighted moment of Eq.~(\ref{eq:CK}) associated with Close--Kumano is nonzero, whereas the ET-type momentum-weighted moment of Eq.~(\ref{eq:LET}) formed from the same data is consistent with zero.  This is not a full-moment test, but it mirrors the theoretical selection of the ET-type weighting in the Bjorken projection.  A direct finite-$Q^2$ test requires the protected combination to be isolated without a model for the unmeasured tensor functions.  The separated $LT$ and $TT$ responses provide this projection.  For inclusive scattering of an unpolarized lepton from a tensor-polarized spin-1 target, the tensor part of the cross section contains
\bea
 \frac{\dd^2\sigma}{\dd x\,\dd Q^2} & = & \mathcal K_\sigma\{\cdots
 +\sqrt{2\epsilon(1+\epsilon)}\,T_{LT}\cos\phi_{LT}\,F_{UT_{LT}}
 +\epsilon\,T_{TT}\cos2\phi_{TT}\,F_{UT_{TT}}\},
 \label{eq:Lcrosssection}
\eea
where $\mathcal K_\sigma$ contains the common kinematic factors.  In the conventions used in current deuteron analyses\footnote{We note the one-to-one correspondence of these response functions with the analytically-continued Budnev's responses (see Appendix~\ref{app:Budnevbasis}), e.g.: $$ F_{UT_{TT}}=-2M|\boldsymbol q|\,\tau_{TT}, \quad F_{UT_{LT}}=-8M|\boldsymbol q|\,\tau_{TL}.$$}~\cite{Cosyn:2020kwu,Cosyn:2024drt},
\begin{subequations}\label{eq:LUTresponses}
\bea
 F_{UT_{LT}} & = & -\frac{M}{3Q}\left[v_\gamma^2 (b_2-3b_4)
 +2(b_2-3b_3)\right],
 \label{eq:LUTresponseLT}\\
 F_{UT_{TT}} & = & -\frac{M}{3\nu}\left(b_2-3b_3\right).
 \label{eq:LUTresponseTT}
\eea
\end{subequations}
The second equation removes the $b_3$ term from the first, such that the inelastic integrand entering the left-hand side of the sum rule of Eq.~(\ref{eq:Linelbalance}) is expressed as
\beq
 b_2-3b_4=
 \frac{3\nu^2}{M(\nu^2+Q^2)}
 \left(2\nu F_{UT_{TT}}-QF_{UT_{LT}}\right).
 \label{eq:Lmeasure}
\eeq
The cancellation is algebraic and requires no assumption about $b_3$.

The geometry in Fig.~\ref{fig:introAlignment} now has a practical meaning.  Let the alignment axis be $\boldsymbol a=(\sin\theta\cos\phi,\sin\theta\sin\phi,\cos\theta)$.  For tensor polarization $P_{zz}=n_++n_--2n_0$, phases can be chosen so that
\begin{subequations}\label{eq:Lgeometry}
\bea
 T_{LT}\cos\phi_{LT} & = & \frac{P_{zz}}{4}\sin2\theta\cos\phi,
 \label{eq:LgeometryLT}\\
 T_{TT}\cos2\phi_{TT} & = & \frac{P_{zz}}{2}\sin^2\theta\cos2\phi.
 \label{eq:LgeometryTT}
\eea
\end{subequations}
Alignment along the virtual photon sets both harmonics to zero.  This is the geometry usually favored for a clean $b_1$ measurement, but it is the wrong geometry for the present sum rule.  A tilted target is needed to obtain $LT$ sensitivity; a transverse setting supplies the strongest $TT$ projection.  An experimental program which combines these settings can therefore reconstruct the protected combination directly.

Although the deuteron is the immediate target of interest, Eqs.~\eqref{eq:Lcrosssection}--\eqref{eq:Lmeasure} follow from the spin-1 response decomposition.  Therefore, any target for which the corresponding tensor harmonics can be prepared and resolved is subject to the same projection.  A complete sum-rule test must combine the elastic endpoint with the full inelastic moment and control the unmeasured small-$x$ region.

\begin{figure}[t]
\centering
\includegraphics[width=0.9\textwidth]{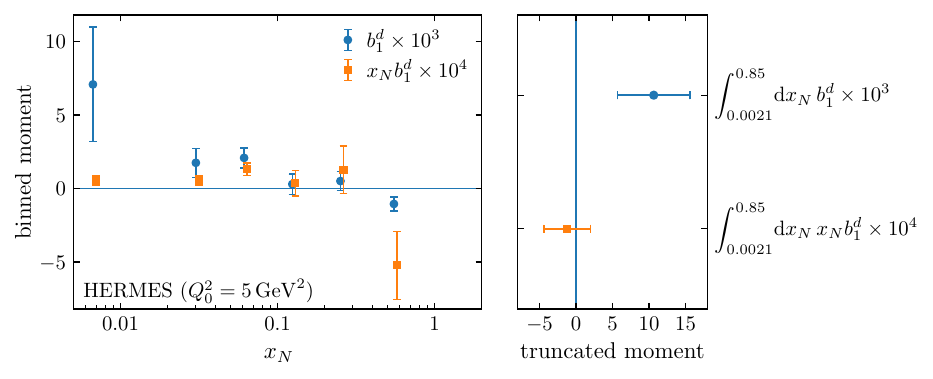}
\caption{Two moments of the deuteron $b_1$ from the HERMES experiment \cite{HERMES:2005pon}. Left: bin contributions. Right: 
combined moments over the measured range, representing Eqs.~\eqref{eq:LHERMES} and \eqref{eq:LHERMESweighted}.}
\label{fig:HERMES}
\end{figure}

The existing HERMES data can nevertheless provide a useful, but deliberately limited, illustration of the Bjorken-limit footprint. Their deuteron analysis used the nucleon-rescaled variable, $x_N\simeq 2x$, and, after evolution to $Q_0^2=5~\mathrm{GeV}^2$, HERMES reported the measured-range first moment~\cite{HERMES:2005pon,Riedl:2005jq}
\beq
 \int_{0.0021}^{0.85}\dd x_N\,b_1^d(x_N,Q_0^2)
 =(10.5\pm3.4_{\rm stat}\pm3.5_{\rm sys})\times10^{-3}.
 \label{eq:LHERMES}
\eeq
 From the same binned data we form instead the momentum-weighted measured-range moment,
\beq
 \int_{0.0021}^{0.85}\dd x_N\,x_N\,b_1^d(x_N,Q_0^2)
 =(-1.2\pm 2.1_{\rm stat}\pm 2.4_{\rm sys})\times10^{-4}.
 \label{eq:LHERMESweighted}
\eeq
The extra power of $x_N$ changes the balance of the bins dramatically, as shown in Fig.~\ref{fig:HERMES}: the HERMES first moment is visibly nonzero over the measured range, whereas our weighted evaluation is consistent with zero.

At least within the measured range, the data follow the same hierarchy selected theoretically in Sec.~\ref{sec:Lope}: the ET momentum-weighted moment is compatible with zero, whereas the Close--Kumano unweighted moment does not need to. Of course for the $x$-weighted integral, the missing data at higher $x$ become more significant. A desirable aim for the JLab experiments would be to fill in that high-$x$ gap and provide a more stringent test of these sum rules.

Obtaining data over a range of $Q^2$ will allow for further theory tests.  The $T-\Delta$ subtraction function \eqref{eq:LPhiTsum}, which is proportional to the exact finite-$Q^2$ weighted moment, provides a concrete example.  The twist-two part of that moment has the target-mass expansion of Eq.~\eqref{eq:xb1TMC}. 
The full physical moment of Eq.~\eqref{eq:LPhiTsum} is not identical to the left-hand side of Eq.~\eqref{eq:xb1TMC}: in addition to the target-mass-corrected twist-two contribution it contains dynamical higher twists and the elastic endpoint, $b_1^\mathrm{el}=\frac{\tau}{2}G_M^2(Q^2)\delta(1-x)$. Measuring the full physical moment of Eq.~\eqref{eq:LPhiTsum} over a range of $Q^2$ can therefore test its approach toward the target-mass-corrected twist-two form, rather than an exactly vanishing moment at finite $Q^2$.

Similar finite-$Q^2$ tests are applicable to the separated $LT$ and $TT$ tensor harmonics of Eq.~\eqref{eq:LUTresponses}, given the OPE-induced prediction in Eq.~\eqref{eq:LWW}.

\FloatBarrier
\section{Conclusion}
\label{sec:Lconclusion}

Equations~\eqref{eq:Lsumrule} and \eqref{eq:LtensorSubtractions} introduce the three independent rank-two sum rules derived from the same Siegert degeneracy, but separated by their dispersive status.  The mixed $TT$--$LT$ pair has opposite crossing parity, so the two
unsubtracted kernels combine there to yield the superconvergent first
moment of $b_2-3b_4$.  The $T\text{--}\Delta$ and $L\text{--}\Delta$ pairs have the same even crossing parity; one subtraction is required, and the corresponding
moments determine two tensor subtraction functions ${\cal M}_{T\text{--}\Delta}(0,Q^2)$ and
${\cal M}_{L\text{--}\Delta}(0,Q^2)$ instead of vanishing.

At large $Q^2$, the superconvergence relation \eqref{eq:Lsumrule} joins smoothly onto the QCD operator expansion.  Detmold's twist-two analysis reproduces the same lowest moment while retaining the kinematic effects of the finite target mass.  The asymptotic moments $\Gamma_n[b_1]$ entering this expansion should not be confused with the exact finite-$Q^2$ moments that determine the subtraction functions; the latter also contain the elastic endpoint and dynamical higher-twist contributions.  In the strict Bjorken projection, and neglecting $b_4$, the relation selects the Efremov--Teryaev type momentum-weighted moment of $b_1$, rather than the unweighted Close--Kumano moment.  HERMES deuteron data make this distinction visible: over the measured range the weighted moment is consistent with zero, whereas the unweighted one is not.  The data mirror the same choice between the two moments that emerges from the theory.  A direct finite-$Q^2$ test requires the separated tensor $LT$ and $TT$ responses, together with the elastic endpoint and control of the unmeasured small-$x$ region.

The three sum rules thus carry the same idea across very different regimes: the high-$Q^2$ OPE picture, the low-$Q^2$ elastic--inelastic balance, and the real-photon tensor dynamics.  Their origin is local; crossing and asymptotics decide how their origin appears in observables.  
Testing them provides interesting opportunities at Jefferson Lab and, in the future, at the EIC.  On the theory side, the superconvergence hypothesis itself deserves scrutiny: a fixed-pole contribution, as found in leading-order QED for the photon--photon analogue, would leave a nonzero right-hand side in Eq.~\eqref{eq:Lsumrule}.  Explicit field-theory evaluations, e.g., one-loop calculations for an elementary massive vector target in massive Yang--Mills QED, provide a near-future test, together with the quadrupole sum rule and the tensor polarizabilities that follow from the low-$Q^2$ expansion.  

\par\medskip
\noindent\textbf{Acknowledgments.} This work is supported by the Deutsche Forschungsgemeinschaft (DFG) through the Collaborative Research Center CRC 1660 ``Hadrons and Nuclei as Discovery Tools'' (Project ID 514321794).  The algebraic derivations were cross-checked, and the scripts generating the figures were written, with the assistance of AI tools, such as Claude (Opus 5.5) and ChatGPT (Sol 5.6); the authors verified all results and take full responsibility for the content of this article.

\small
\par\medskip
%\noindent{\large\bfseries Supplementary Material.} The appendices collect conventions and derivations that would interrupt the main argument. Results quoted in the main text are repeated only where needed to make a derivation self-contained.

\appendix
\section{Conventions and response dictionaries}
\label{app:dictionary}

This appendix fixes the kinematic and polarization conventions and gives the translations among the HJM covariant basis, the Budnev helicity-response basis, and the tensor basis of Edelmann, Piller, and Weise. We use $p^2=M^2$, $q^2=-Q^2$, $\nu=p\!\cdot q/M$, $\boldsymbol q^{\,2}=\nu^2+Q^2$, $v_\gamma^2=\boldsymbol q^{\,2}/\nu^2$, and $\gamma=(1-v_\gamma^2)^{-1/2}=\nu/(iQ)$.

The massive spin-1 target polarization vector is $\chi^\mu$, with $p\!\cdot\chi=0$ and $\chi^*\!\cdot\chi=-1$; $\epsilon_\lambda^\mu$ is reserved for photon polarization vectors.

\subsection{Covariant HJM basis}
\label{app:HJMbasis}

To make current conservation manifest, we define the $q$-transverse quantities
\begin{subequations}\label{eq:HJMtransverse}
\bea
 \widetilde g_{\mu\nu} & = & g_{\mu\nu}-\frac{q_\mu q_\nu}{q^2},
 \label{eq:HJMtransverseMetric}\\
 \widetilde p_\mu & = & p_\mu-\frac{M\nu}{q^2}q_\mu,\qquad
 \widetilde\chi_\mu=\chi_\mu-\frac{q\!\cdot\chi}{q^2}q_\mu,
 \label{eq:HJMtransverseVectors}
\eea
\end{subequations}
and the tensor (rank-two) polarization of the target projected onto the photon direction, with the spin average subtracted:
\beq
 A_\chi\equiv
 (q\!\cdot\chi^*)(q\!\cdot\chi)-\frac{\boldsymbol q^{\,2}}{3}.
 \label{eq:HJMAlignScalar}
\eeq
The four symmetric rank-two HJM tensors can then be written as
\begin{subequations}\label{eq:HJMstructures}
\bea
 r_{\mu\nu} & = & \frac{A_\chi}{\nu^2}\,\widetilde g_{\mu\nu},\label{eq:HJM_r}\\
 s_{\mu\nu} & = & \frac{2A_\chi}{M\nu^3}\,
 \widetilde p_\mu\widetilde p_\nu,\label{eq:HJM_s}\\
 t_{\mu\nu} & = & \frac{1}{2\nu^2}\Big\{
 (q\!\cdot\chi^*)
 [\widetilde p_\mu\widetilde\chi_\nu+\widetilde p_\nu\widetilde\chi_\mu]
 +(q\!\cdot\chi)
 [\widetilde p_\mu\widetilde\chi^*_\nu+\widetilde p_\nu\widetilde\chi^*_\mu]
 -\frac{4\nu}{3M}\widetilde p_\mu\widetilde p_\nu\Big\},\label{eq:HJM_t}\\
 u_{\mu\nu} & = & \frac{M}{\nu}\Big[
 \widetilde\chi^*_\mu\widetilde\chi_\nu
 +\widetilde\chi^*_\nu\widetilde\chi_\mu
 +\frac23\widetilde g_{\mu\nu}
 -\frac{2}{3M^2}\widetilde p_\mu\widetilde p_\nu\Big].\label{eq:HJM_u}
\eea
\end{subequations}
 The complete absorptive forward VVCS tensor is written as:
\bea
 W_{\mu\nu} & = & -F_1\widetilde g_{\mu\nu}
 +\frac{F_2}{M\nu}\widetilde p_\mu\widetilde p_\nu\nonumber\\
 &&+\frac{i g_1}{\nu}\epsilon_{\mu\nu\alpha\beta}q^\alpha s^\beta
 +\frac{i g_2}{\nu^2}\epsilon_{\mu\nu\alpha\beta}q^\alpha
 \left(\nu s^\beta-\frac{s\!\cdot q}{M}p^\beta\right)\nonumber\\
 &&-b_1 r_{\mu\nu}
 +\frac16 b_2(s_{\mu\nu}+t_{\mu\nu}+u_{\mu\nu})
 +\frac12 b_3(s_{\mu\nu}-u_{\mu\nu})
 +\frac12 b_4(s_{\mu\nu}-t_{\mu\nu}), 
 \label{eq:HJMcovariant}
\eea
where we introduce the dimensionless spin vector
\beq
 s^\mu=-\frac{i}{M}\,\epsilon^{\mu\alpha\beta\rho}
 \chi^*_{\alpha}\chi_{\beta}p_{\rho}.
 \label{eq:HJMspinVector}
\eeq
This is the HJM decomposition in the normalization used in the main text~\cite{Hoodbhoy:1988am}. 

\subsection{Dictionary to the Budnev light-by-light response basis}
\label{app:Budnevbasis}

The HJM decomposition can also be related to the older basis
of Budnev \emph{et al.}~\cite{Budnev:1975poe}, used for light-by-light (LbL)
scattering. The connection is obtained by analytic continuation
from a spacelike virtual photon treated as the target to a massive spin-1 target.

Two independent convention choices require care. First, Budnev's superscript $a$ labels the antisymmetric Compton tensor and hence the spin-dependent $g_{1,2}$ sector; it does \emph{not} denote odd crossing parity~\cite{Budnev:1975poe}. Crossing under $\nu\to-\nu$ is a separate classification. The spin sector contains a crossing-odd $TT$ and a crossing-even mixed $TL$ response, whereas the rank-two sector contains a crossing-even $TT$ and a crossing-odd mixed $TL$ response. This opposite assignment is the origin of the different explicit crossing factors in the BC and tensor combinations.

Second, Budnev's spacelike longitudinal state carries an imaginary phase, while later LbL analyses often use a real longitudinal polarization~\cite{Pascalutsa:2012pr,Gerardin:2017ryf}. The two choices are related by rephasing the external state, which changes the relative sign of raw mixed-helicity amplitudes and interchanges the names attached to the sum and difference of $W_{++,00}$ and $W_{0+,-0}$ in parts of the literature.

We retain the Budnev sector labels and crossing assignments, as in Refs.~\cite{Pascalutsa:2024hvy,Pascalutsa:2025suw}, but use the standard real longitudinal convention of Refs.~\cite{Pascalutsa:2012pr,Gerardin:2017ryf}. The corresponding rephasing is applied at the helicity-amplitude level before the two bases are compared. Physical HJM relations are unchanged when one convention is used consistently. For compactness we set $M=1$; factors of $M$ can be restored dimensionally. In terms of the Budnev responses,
\begin{subequations}\label{eq:BudnevHelicityDefs}
\bea
 W_{++,++}+W_{+-,+-} & = & 4|\boldsymbol q|\,\sigma_{TT},\qquad W_{++,++}-W_{+-,+-}=4|\boldsymbol q|\,\tau_{TT}^{a},\\
 W_{++,--} & = & 2|\boldsymbol q|\,\tau_{TT},\qquad W_{00,00}=2|\boldsymbol q|\,\sigma_{LL},\\
 W_{+0,+0} & = & 2|\boldsymbol q|\,\sigma_{TL},\qquad W_{0+,0+}=2|\boldsymbol q|\,\sigma_{LT},\\
 W_{++,00}+W_{0+,-0} & = & 4|\boldsymbol q|\,\tau_{TL},\qquad W_{++,00}-W_{0+,-0}=4|\boldsymbol q|\,\tau_{TL}^{a}.
\eea
\end{subequations}
Deriving the same helicity projections from the HJM basis, we obtain
the following dictionary:
\begin{subequations}\label{eq:HJMtoBudnev}
\bea
 2|\boldsymbol q|\,\sigma_{TT}
  & = & F_1-\frac{v_\gamma^2}{3}b_1+\frac{1}{6\nu}\left(\frac13b_2-b_3\right),\label{eq:HJMtoBudnevTT}\\
 2|\boldsymbol q|\,\sigma_{TL}
  & = & F_1+\frac{2v_\gamma^2}{3}b_1-\frac{1}{3\nu}\left(\frac13b_2-b_3\right),\\
 2|\boldsymbol q|\,\sigma_{LT}
  & = & -F_1+\frac{\nu v_\gamma^2}{Q^2}F_2
 +\frac{v_\gamma^2}{3}b_1
 -\frac{\nu(v_\gamma^4+v_\gamma^2+1)}{9Q^2}b_2\nonumber\\
 &&\hspace{1.6cm}-\frac{1+v_\gamma^2}{3\nu}b_3
 -\frac{v_\gamma^2}{3\nu}b_4,\\
 2|\boldsymbol q|\,\sigma_{LL}
  & = & -F_1+\frac{\nu v_\gamma^2}{Q^2}F_2
 -\frac{2v_\gamma^2}{3}b_1
 +\frac{2\nu(v_\gamma^4+v_\gamma^2+1)}{9Q^2}b_2\nonumber\\
 &&\hspace{1.6cm}+\frac{2(1+v_\gamma^2)}{3\nu}b_3
 +\frac{2v_\gamma^2}{3\nu}b_4,\\
 2|\boldsymbol q|\,\tau_{TT} & = & \frac{1}{\nu}\left(\frac13b_2-b_3\right),\label{eq:HJMtoBudnevTauTT}\\
 4|\boldsymbol q|\,\tau_{TL}
  & = & \frac{1}{Q}\left[\frac13b_2-b_3-\frac12v_\gamma^2\left(b_4-\frac13b_2\right)\right],\label{eq:HJMtoBudnevTauTL}\\
 4|\boldsymbol q|\,\tau_{TT}^{a} & = & 2\left(g_1-\frac{Q^2}{\nu^2}g_2\right),\label{eq:HJMtoBudnevTauTTa}\\
 4|\boldsymbol q|\,\tau_{TL}^{a} & = & -\frac{2Q}{\nu}(g_1+g_2).\label{eq:HJMtoBudnevTauTLa}
\eea
\end{subequations}
As a check, averaging the diagonal entries over the three target substates cancels every $b_i$ term and leaves the unpolarized responses. The tensor-interference entries reproduce Eq.~\eqref{eq:Lresponses}, while their crossing assignments are manifest: $\tau_{TT}^{a}$ is odd and $\tau_{TL}^{a}$ even in the spin sector, whereas $\tau_{TT}$ is even and $\tau_{TL}$ odd in the rank-two sector. At $\boldsymbol q^{\,2}=0$, the coefficients of $F_2$, $b_1$, and $b_4$ disappear, leaving only $F_1$, $\frac13b_2-b_3$, and $g_1+g_2$. This is the covariant form of the same degeneracy used in Sec.~\ref{sec:Lsiegert}.

\subsection{Dictionary to the Edelmann--Piller--Weise tensor basis}
\label{app:Edelmann}

The Edelmann--Piller--Weise small-$x$ basis is organized directly in helicity responses~\cite{Edelmann:1997qe,Edelmann:1997ik}. In the rank-two subspace, let $\Delta$ denote the double-helicity-flip response and $b_3^E$ the mixed tensor response. Setting $M=1$, with $v_\gamma=|\boldsymbol q|/\nu$ and the normalization matched to Eqs.~\eqref{eq:BudnevHelicityDefs}--\eqref{eq:HJMtoBudnev}, gives
\begin{subequations}\label{eq:HJMtoEdelmann}
\bea
 \Delta & = & W_{++,--}=\frac{1}{\nu}\left(\frac13b_2-b_3\right),\label{eq:HJMtoEdelmannDelta}\\
 b_3^E & = & \frac{\nu}{2Qv_\gamma}\left(W_{++,00}+W_{0+,-0}\right)\nonumber\\
  & = & \frac{1}{2\nu v_\gamma(v_\gamma^2-1)}
 \left[\frac13b_2-b_3-\frac12v_\gamma^2\left(b_4-\frac13b_2\right)\right].\label{eq:HJMtoEdelmannB3}
\eea
\end{subequations}
Equivalently, the inverse transformation needed for the present sum rules can be written as
\begin{subequations}\label{eq:EdelmannInv}
\bea
 \frac13b_2-b_3 & = & \nu\,\Delta,
 \label{eq:EdelmannInvA}\\
 b_4-\frac13b_2
  & = & \frac{2\nu}{v_\gamma^2}\,\Delta
 -\frac{4\nu(v_\gamma^2-1)}{v_\gamma}\,b_3^E,
 \label{eq:EdelmannInvB}\\
 b_2-3b_4
  & = & -\frac{6\nu}{v_\gamma^2}\,\Delta
 +\frac{12\nu(v_\gamma^2-1)}{v_\gamma}\,b_3^E,
 \label{eq:EdelmannInvProtected}\\
 \frac13b_2+b_3+b_4
  & = & b_2+\nu\left(\frac{2}{v_\gamma^2}-1\right)\Delta
 -\frac{4\nu(v_\gamma^2-1)}{v_\gamma}\,b_3^E.
 \label{eq:EdelmannInvLDelta}
\eea
\end{subequations}
The last two lines display explicitly the tensor superconvergence relation and the $L-\Delta$ subtraction-function sum rule.   The inverse factors of $v_\gamma$ and $v_\gamma^2$ also show why individual HJM coefficients need not have a smooth term-by-term Bjorken limit.  The limit $v_\gamma\to1$ must therefore be taken at the level of physical helicity responses or of the complete HJM combinations, not term by term in the basis transformation. 

\section{Elastic HJM structure functions}
\label{app:elastic}

The main text gives the elastic contributions of all three tensor sum rules and retains only the real-photon consequence of the elastic--inelastic balance.  Here we list the complete elastic HJM structure functions from which these contributions follow, using the standard spin-one form-factor convention~\cite{Arnold:1979cg,JLABt20:2000uor}:
\begin{subequations}\label{eq:elasticAll}
\bea
 F_1^{\rm el} & = & \frac{1+\tau}{3}G_M^2\delta(1-x),
 \label{eq:elasticF1}\\
 F_2^{\rm el} & = & \left(G_C^2+\frac23\tau G_M^2+\frac89\tau^2G_Q^2\right)\delta(1-x),
 \label{eq:elasticF2}\\
 g_1^{\rm el} & = & \frac12G_M\left(G_C+\frac\tau2G_M+\frac\tau3G_Q\right)\delta(1-x),
 \label{eq:elasticg1}\\
 g_2^{\rm el} & = & \frac\tau2G_M\left(G_C-\frac12G_M+\frac\tau3G_Q\right)\delta(1-x),
 \label{eq:elasticg2}\\
 b_1^{\rm el} & = & \frac\tau2G_M^2\delta(1-x),
 \label{eq:elasticb1}\\
 b_2^{\rm el} & = & \left\{\tau G_M^2+\frac{4\tau^2}{1+\tau}G_Q
 \left(G_C-G_M+\frac\tau3G_Q\right)\right\}\delta(1-x),
 \label{eq:elasticb2}\\
 b_3^{\rm el} & = & \left\{-\frac{\tau(2+3\tau)}3G_M^2
 +\frac{4\tau^2}{9(1+\tau)}G_Q[3(G_C-G_M)+\tau G_Q]\right\}\delta(1-x),
 \label{eq:elasticb3}\\
 b_4^{\rm el} & = & \left\{\frac{\tau(1+6\tau)}3G_M^2
 +\frac{4\tau^2}{9(1+\tau)}G_Q[3G_C+3(2+3\tau)G_M+\tau G_Q]\right\}\delta(1-x).
 \label{eq:elasticb4}
\eea
\end{subequations}

\setlength{\bibsep}{1.2pt plus 0.3pt}
\renewcommand{\bibfont}{\small}
%\section*{References}
\bibliography{refs}
\end{document}